\documentclass[pdflatex,sn-nature]{sn-jnl}

\usepackage{graphicx}
\usepackage{multirow}
\usepackage{amsmath,amssymb,amsfonts}
\usepackage{amsthm}
\usepackage[title]{appendix}
\usepackage{xcolor}
\usepackage{textcomp}
\usepackage{manyfoot}
\usepackage{booktabs}
\usepackage{algorithm}
\usepackage{hyperref}
\usepackage{algorithmicx}
\usepackage{algpseudocode}
\usepackage{listings}
\usepackage[acronym,nomain]{glossaries}

\newif\ifshowchanges
\showchangesfalse
\ifshowchanges
    
\else
    
\fi

\newacronym{ece}{ECE}{electron cyclotron emission}
\newacronym{bes}{BES}{beam emission spectroscopy}
\newacronym{co2}{CO\textsubscript{2}}{carbon dioxide}
\newacronym{eccd}{ECCD}{electron cyclotron current drive}
\newacronym{ae}{AE}{Alfv\'en eigenmode}
\newacronym{tm}{TM}{tearing mode}
\newacronym{elm}{ELM}{edge-localized mode}

\newacronym{stft}{STFT}{short-time Fourier transform}
\newacronym{cps}{CPS}{cross-power spectrum}
\newacronym{cdf}{CDF}{cumulative distribution function}
\newacronym{eeg}{EEG}{electroencephalography}
\newacronym{tv}{TV}{total variation}

\newacronym{dclde}{DCLDE}{Detection, Classification, Localization, and Density Estimation}
\newacronym{raddet}{RadDet}{Radar Spectrum Detection Dataset}

\newacronym{mse}{MSE}{mean squared error}
\newacronym{gpu}{GPU}{graphics processing unit}
\newacronym{cpu}{CPU}{central processing unit}
\newacronym{faith}{FAITH}{Fusion Artificial Intelligence and Toolkit Hub}

\glsdisablehyper

\theoremstyle{thmstyleone}

\theoremstyle{thmstyletwo}

\theoremstyle{thmstylethree}

\begin{document}

\title[TokEye]{A general extractor for coherent patterns across time-frequency measurements}

\author*[1]{\fnm{Nathaniel} \sur{Chen}}\email{nathaniel@princeton.edu}
\author[1]{\fnm{Kouroche} \sur{Bouchiat}}
\author[1]{\fnm{Peter} \sur{Steiner}}
\author[1]{\fnm{Andrew} \sur{Rothstein}}
\author[2]{\fnm{David} \sur{Smith}}
\author[3]{\fnm{Max} \sur{Austin}}
\author[3]{\fnm{Michael} \spfx{van} \sur{Zeeland}}
\author[4,5]{\fnm{Marie} \sur{Roch}}
\author[1]{\fnm{Azarakhsh} \sur{Jalalvand}}
\author*[1,6]{\fnm{Egemen} \sur{Kolemen}}\email{ekolemen@princeton.edu}

\affil*[1]{\orgdiv{Mechanical \& Aerospace Engineering}, \orgname{Princeton University}, \orgaddress{\street{41 Olden Street}, \city{Princeton}, \postcode{08544}, \state{NJ}, \country{USA}}}
\affil[2]{\orgdiv{Department of Engineering Physics}, \orgname{University of Wisconsin-Madison}, \orgaddress{\city{Madison}, \postcode{53706}, \state{WI}, \country{USA}}}
\affil[3]{\orgname{DIII-D National Fusion Facility}, \orgaddress{\city{San Diego}, \postcode{92186}, \state{CA}, \country{USA}}}
\affil[4]{\orgname{San Diego State University}, \orgaddress{\city{San Diego}, \postcode{92182}, \state{CA}, \country{USA}}}
\affil[5]{\orgname{Scripps Institution of Oceanography}, \orgaddress{\city{La Jolla}, \postcode{92037}, \state{CA}, \country{USA}}}
\affil[6]{\orgname{Princeton Plasma Physics Laboratory}, \orgaddress{\city{Princeton}, \postcode{08540}, \state{NJ}, \country{USA}}}

\abstract{Scientific discovery increasingly depends on finding meaningful patterns within volumes of sensor data too large for manual inspection. Existing time-frequency event extractors are often engineered or trained on manually labeled examples for each instrument, which can be laborious and difficult to adapt to new environments. To address this, we introduce TokEye, a reference-free method that learns a general criterion for coherent structure from heterogeneous fusion measurements. A model developed under this method, without retraining, can extract signals from various other fields including marine bioacoustics and wideband radar. It recovers 82.5\% of expert-annotated stellarator signals, detects 78\% of dolphin whistles, and identifies radar events with 81.5\% average precision across six benchmarks, while processing full experimental spectrograms in under one second on a GPU. This transfer across unrelated sensors and physical systems reveals common exploitable structure in time-frequency measurements and establishes a scalable foundation for automated scientific signal discovery.}

\maketitle

\section{Introduction}\label{sec:introduction}
Across many measurement-driven sciences, analysis begins by isolating sparse, coherent structure from noisy time-frequency measurements. Much of this process still relies on hand-tuned filters or supervised models whose labels encode instrument-specific expert judgment. These approaches must be recreated for each sensor and can favor recognized event classes over unexpected but physically meaningful structure. We therefore ask whether a useful automated extraction criterion can generalize across instruments, facilities, and scientific domains without clean references or target-domain retraining.

We first test transfer from DIII-D to the TJ-II stellarator, another magnetically confined fusion device. TJ-II's non-axisymmetric magnetic configuration differs fundamentally from the axisymmetric DIII-D tokamak and is equipped with a different diagnostic suite \cite{alejaldreFirstPlasmasTJII1999}. This comparison asks whether the learned criterion can generalize to magnetic confinement reactors at any facility facility. We then test the broader cross-domain question using two established benchmarks: \gls{dclde} marine-mammal recordings, which contain dolphin whistles obscured by snapping shrimp and other background noise, and \gls{raddet}, a wideband-radar spectrogram dataset spanning diverse signal shapes, densities, and noise conditions \cite{rochAutomatedExtractionOdontocete2011, rochDCLDE2011Data2025, huangRadDetWidebandDataset2025a}.

Contribution: TokEye addresses these questions through an optimization-driven process that combines broadband baseline correction with complex-valued, cross-channel self-supervised denoising, using the former to expose coherent residual structure to the latter. Under linear Gaussian assumptions, the conditional estimator reduces to the classical form determined by auto- and cross-spectral densities, positioning the learned estimator as a nonlinear generalization of cross-power analysis. The multistage process is then distilled into a fast single-channel surrogate trained on DIII-D; its weights are held fixed when applied to a second confinement device, marine bioacoustics, and radar. This design tests whether coherent structure can serve as a transferable measurement target across otherwise unrelated sensors and physical systems, offering a general framework for automated scientific signal discovery in time-frequency measurements.

\section{Results}\label{sec:results}
\begin{figure}[t]
    \centering
    \includegraphics[width=\textwidth]{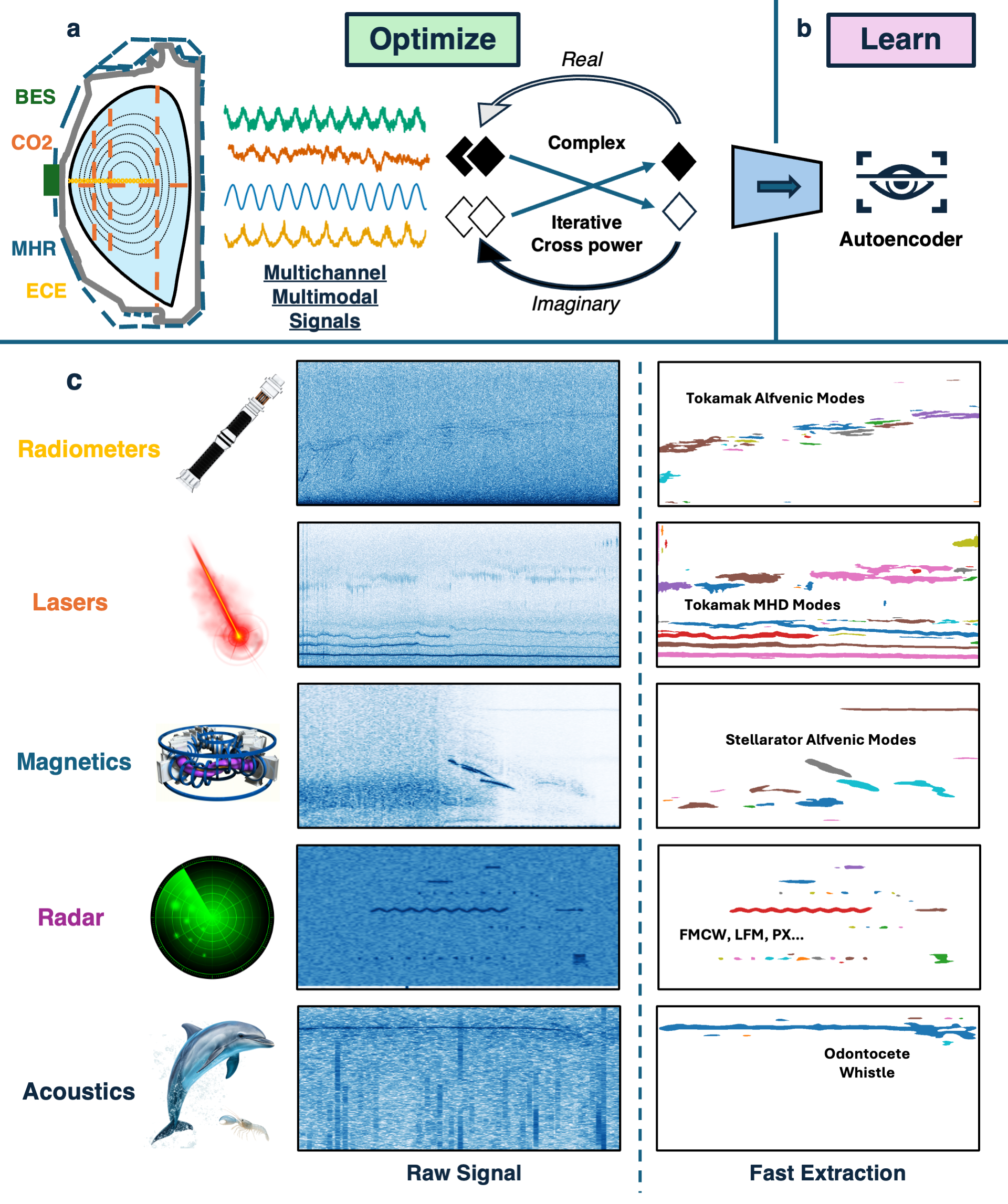}
    \caption{\textbf{Preparation and usage of Tokeye.} (a) Multichannel, multimodal DIII-D diagnostics (\gls{bes}, \gls{co2} interferometry, magnetics, and \gls{ece}) provide heterogeneous source measurements. A broadband baseline correction precedes complex cross-channel self-supervised denoising, allowing the denoiser to learn coherent residual structure. (b) The process is distilled into a single-sensor surrogate. (c) Zero-shot evaluation on a diverse set of DIII-D tokamak diagnostics, TJ-II stellarator diagnostics, wideband-radar data, and marine bioacoustic recordings.}
    \label{fig:overview}
\end{figure}

TokEye tests a source-to-target transfer design (Fig.~\ref{fig:overview}). Heterogeneous diagnostics from the DIII-D tokamak provide the source data for a four-stage, reference-free training process: complex time-frequency transformation, broadband baseline removal, multichannel self-supervised denoising, and knee-point thresholding; this is then distilled onto a fast single-channel model. The resulting model is evaluated qualitatively on  DIII-D, and compared to benchmark datasets including TJ-II, \gls{dclde}, and RADDET. Inference requires neither multiple channels nor the original denoiser.

\begin{figure}[t]
    \centering
    \includegraphics[width=\textwidth]{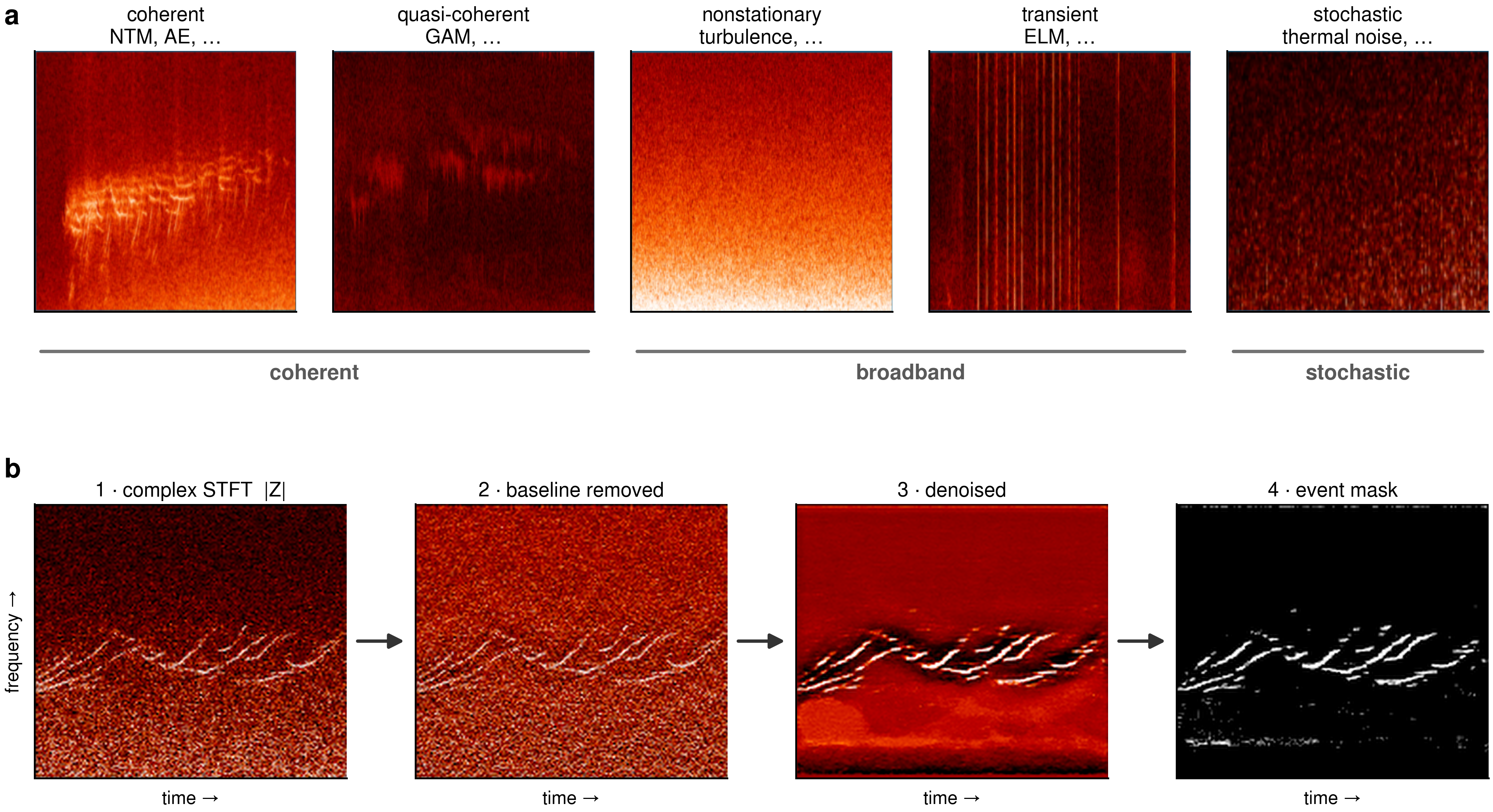}
    \caption{\textbf{Signal taxonomy and processing pipeline.} (a)~Signal taxonomy with example spectra, grouped into coherent, broadband, and stochastic supercategories. (b)~Processing pipeline example: complex \gls{stft}, baseline removal, multichannel denoising, and thresholding.}
    \label{fig:concept}
\end{figure}

\subsection{Cross-device transfer to the TJ-II stellarator}\label{sec:results-tjii}

The pretrained model is applied to the TJ-II stellarator \cite{alejaldreFirstPlasmasTJII1999}. At the fixed default threshold of 0.5, the model automatically recovers 82.5\% of the expert-annotated mode bins. Its boundary-tolerant PR-AUC reaches 0.82 at a $\pm2$-bin tolerance (Section~\ref{sec:results-fp}). For context, the supervised CNN for which this dataset was introduced reported 57.9\% mode-pixel recall \cite{bustosAutomaticIdentificationMHD2021}. Operating points are reported in Supplementary Table~\ref{tab:tjii_vs_bustos}.

The closest label-free in-domain method on TJ-II is the Elastic Random Mode Decomposition (ERMD) \cite{zapata-cornejoNovelUnsupervisedMachine2024}, which clusters gabor atoms of raw Mirnov-coil signals. A direct numerical comparison is not possible since its output is a set of clusters rather than binary masks. The methods also differ in scope and portability. ERMD targets Alfv\'enic activity specifically whereas the pipeline here extracts general coherent fluctuations.

A leave-one-out ablation on this same TJ-II~2021 set, using a reduced training pipeline, isolates the contribution of each design choice-multichannel denoising, baseline removal, and the complex representation (Supplementary Methods, Supplementary Tables~\ref{tab:ablation} and~\ref{tab:ablation_stats}).

\begin{figure}[t]
    \centering
    \includegraphics[width=\textwidth]{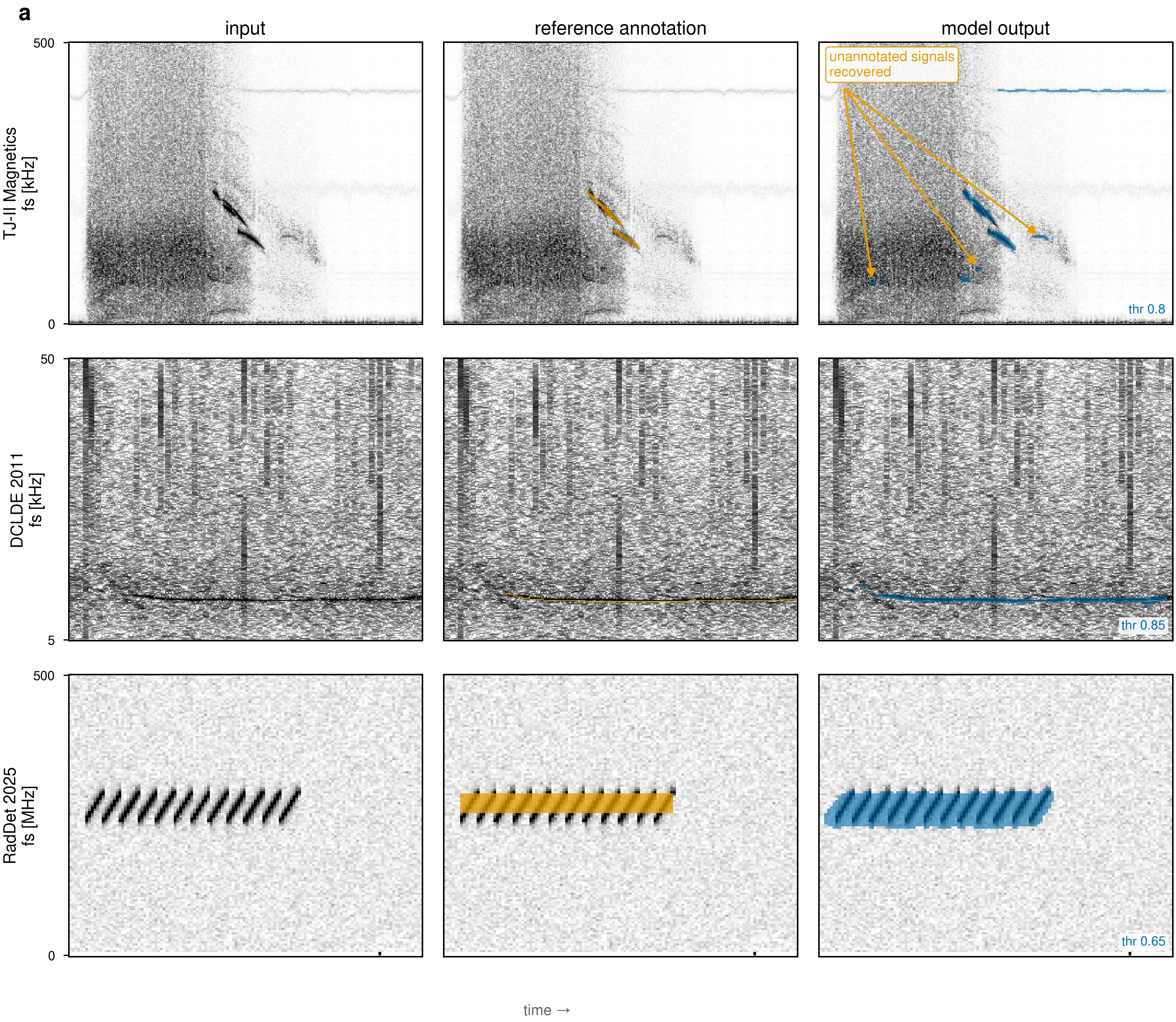}
    \caption{\textbf{Zero-shot transfer of the single DIII-D-trained model.} Rows: TJ-II stellarator magnetics, a \glsentryshort{dclde} 2011 dolphin whistle, and a \glsentryshort{raddet} radar pulse. Columns: input, reference annotation, and model output at the reported F1 operating threshold. Arrows mark coherent structure recovered beyond the reference annotation.}
    \label{fig:transfer}
\end{figure}

\subsection{Cross-domain transfer to marine bioacoustic vocalizations}\label{sec:results-dclde}

We evaluate the same pretrained model on the \gls{dclde} 2011 marine-mammal vocalization benchmark, which provides annotated dolphin tonal whistles amid broadband interference from snapping shrimp, odontocete echolocation clicks, and ambient noise \cite{rochAutomatedExtractionOdontocete2011, rochDCLDE2011Data2025}. At the default operating threshold (0.5) the pipeline recovers 78\% of annotated whistles under an any-overlap criterion (Supplementary Table~\ref{tab:dclde_event}). Pixel metrics on the same predictions are recall $0.72$ at the 0.5 threshold and, PR-AUC $0.73$ with a $\pm2$-bin tolerance (Supplementary Table~\ref{tab:dclde_per_species}). Similar to TJ-II, we find that many of the false positives are narrowband, with only $\sim$5\% broadband (Supplementary Table~\ref{tab:dclde_fp_morphology}, Extended Data Fig.~\ref{fig:dclde_failure}). Pixel-metrics also be dramatically improved when boundary tolerance is increased by a small amount (Section~\ref{sec:results-fp}).

\subsection{Cross-domain transfer to radar detection}\label{sec:results-raddet}

The same pretrained model is evaluated on \gls{raddet}, a public benchmark for radar pulse detection and modulation classification \cite{huangRadDetWidebandDataset2025a}. Across all six dataset variants, mean event average precision is 0.815. At the validation-selected operating points, mean precision is 0.913, recall is 0.841, and F1 is 0.874 (Supplementary Table~\ref{tab:raddet_per_variant}).

\subsection{Pixel precision understates detection}\label{sec:results-fp}

\begin{figure}[t]
    \centering
    \includegraphics[width=\textwidth]{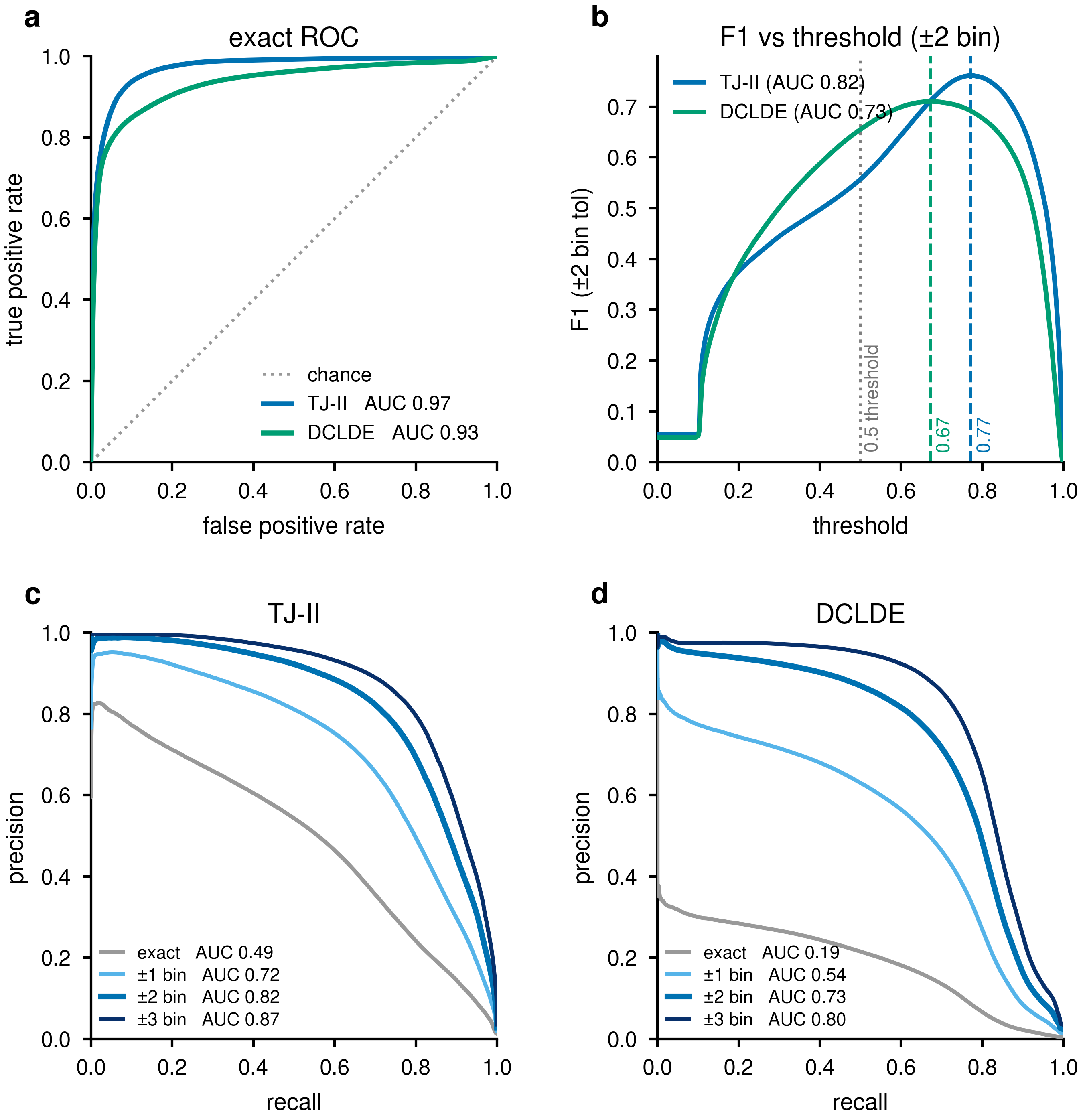}
    \caption{\textbf{Threshold-independent detection and the effect of boundary tolerance.} Bin-level evaluation of the DIII-D-trained model on TJ-II and \gls{dclde}. (a)~ROC-AUC is 0.97 on TJ-II and 0.93 on \gls{dclde}. (b)~F1 as a function of detection threshold at a $\pm2$-bin tolerance; the dotted line marks the fixed default threshold (0.5), and the dashed lines mark F1-maximizing thresholds selected from the annotated evaluation data. (c,~d)~PR curves at increasing boundary tolerance for TJ-II~(c) and \gls{dclde}~(d). PR-AUC rises from 0.49 and 0.19 at exact-bin matching (gray) to 0.82 and 0.73 at a $\pm2$-bin tolerance.}
    \label{fig:pr_curves}
\end{figure}

The bin-level precision reported above is measured at exact spatial agreement, a stringent criterion for time-frequency segmentation. The freedom an annotator has in drawing a mode boundary may place predictions a bin or two from the reference even when the same event is detected (Extended Data Fig.~\ref{fig:dclde_failure}). Therefore boundary-aware segmentation evaluation handles this by scoring matches within a small tolerance band rather than at exact overlap \cite{csurkaWhatGoodEvaluation2013}. We summarize threshold sensitivity by the area under the precision-recall curve (PR-AUC, equivalently the average precision), computed at exact match and at tolerances of $\pm1$, $\pm2$, and $\pm3$ bins (Fig.~\ref{fig:pr_curves}). \gls{raddet} was not included in this analysis since it was designed for instance detection and is multiscale, making consistent pixel-boundary evaluations non-trivial. On TJ-II the bin-level PR-AUC quickly rises from 0.49 at exact match to 0.72, 0.82, and 0.87 at $\pm1$, $\pm2$, and $\pm3$ bins; on \gls{dclde} it jumps from 0.19 to 0.54, 0.73, and 0.80. This bin-level PR-AUC is distinct from the contour-based overlap average precision of Section~\ref{sec:results-dclde}. While the receiver-operating-characteristic (ROC) provides an exceptional value (0.97 on TJ-II, 0.93 on \gls{dclde}, Fig.~\ref{fig:pr_curves}a), the annotated mode bins are a small minority of all time-frequency bins. This is the reason we summarize with PR-AUC.

The large gain from one to two bin tolerance shows that small localization differences account for a substantial part of the exact-bin disagreement. The remaining nominal false-positive bins also have signal-like intensity distributions: on TJ-II, their mean spectrogram intensity is 2.40, compared with 2.90 for true-positive bins and $-0.03$ for confirmed background, and their Wasserstein distance is 0.50 from the true-positive distribution versus 2.43 from the background \cite{villaniOptimalTransport2009a} (Extended Data Fig.~\ref{fig:fp_intensity}). These observations are consistent with boundary mismatch or incomplete annotations but do not exclude structured clutter.

\subsection{Extracting events across DIII-D fluctuation diagnostics}\label{sec:results-d3d}

We examine the model's output on held-out observations from the four DIII-D diagnostics it was trained on: magnetics, \gls{ece} radiometry, \gls{co2} density interferometry, and \gls{bes} (Fig.~\ref{fig:diiid}). The surrogate's second output channel, which isolates broadband transients, is shown for these shots in Extended Data Fig.~\ref{fig:diiid_transient}.

\subsubsection{Case study: fluctuation signatures during tearing-mode suppression}

\begin{figure}[t]
    \centering
    \includegraphics[width=\textwidth]{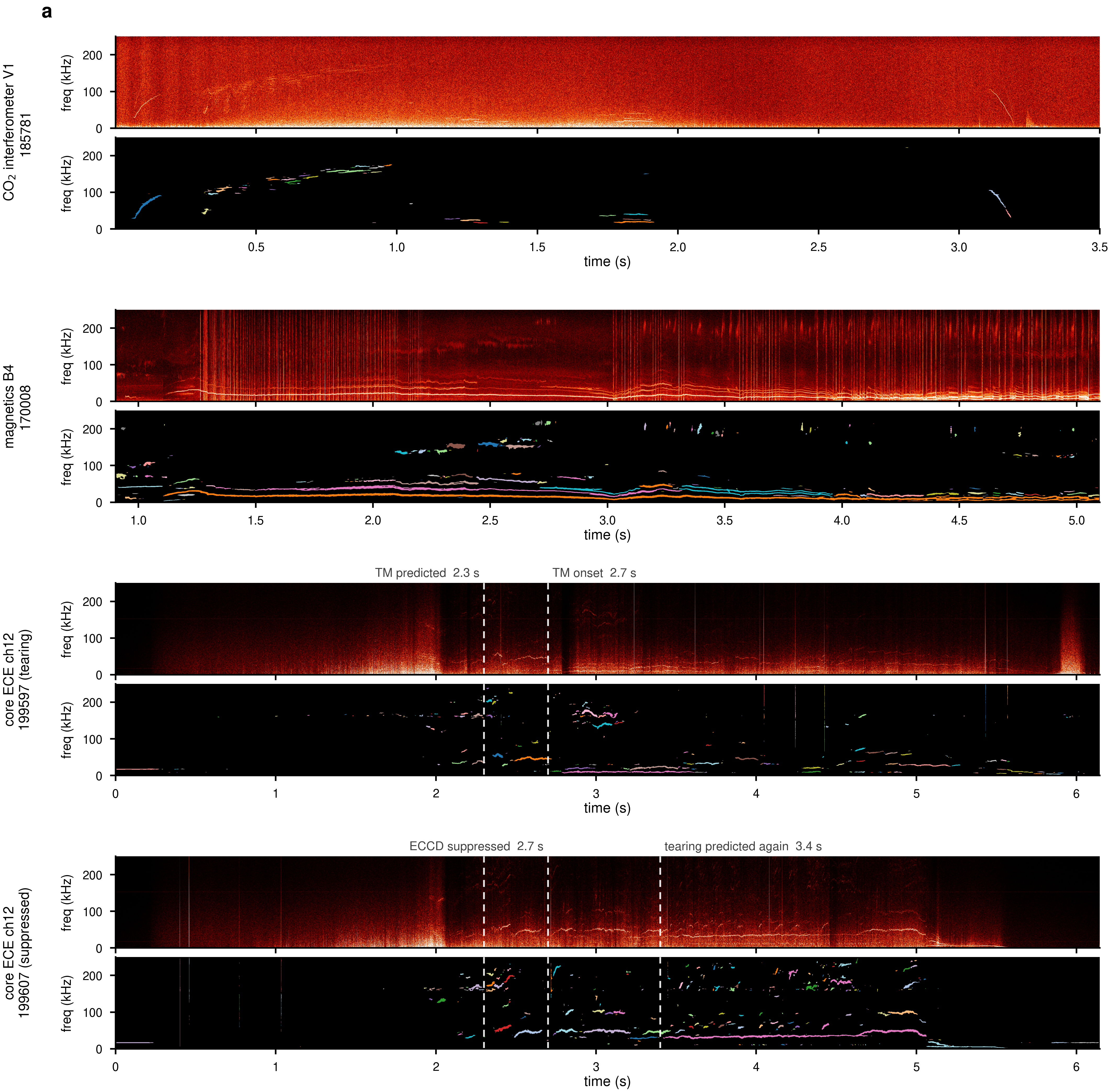}
    \caption{\textbf{Event extraction across DIII-D diagnostics.} \gls{co2} interferometer (shot 185781), high frequency magnetics (shot 170008), and core \gls{ece} for the tearing-unstable and \gls{eccd}-suppressed pair (199597 and 199607). Each pair shows the input spectrogram above the extracted per-event masks. Dashed lines mark tearing-mode timings: ``TM predicted 2.3\,s'' is the onset predicted by the model in Farr\'e-Kaga et~al.\ \cite{farre-kagaInterpretingAIFusion2026}, where TokEye begins to flag coherent activity starting 2.0\,s.}
    \label{fig:diiid}
\end{figure}

\gls{ece} radiometry from DIII-D shots 199597 and 199607 provides a case study illustrating how the model output can facilitate automatic discovery of physics. Both shots come from a tearing-mode control campaign in which shot 199607 applied \gls{eccd} to suppress the tearing instability observed in shot 199597 \cite{farre-kagaInterpretingAIFusion2026}. In the tearing-unstable shot, the pipeline flags mode activity at 2.0\,s in the core, 0.7\,s before its onset at 2.7\,s, and in 199607 it flags renewed tearing activity at 3.4\,s following \gls{eccd} suppression (Fig.~\ref{fig:diiid}). The correspondence is consistent with the Shapley-value attribution reported for the same dataset in Farr\'e-Kaga et~al.\ \cite{farre-kagaInterpretingAIFusion2026}, where core \gls{ece} channels carry the strongest predictive signal for tearing-mode onset. We note that the spectrograms here are complementary to the prediction, which derives from electron-temperature profiles. Edge \gls{ece} channels processed by the pipeline do not exhibit the same activity.

\subsection{Per-domain summary and inference cost}\label{sec:results-performance}

At inference, a full DIII-D shot forms a single-channel spectrogram of 512 frequency bins by roughly 10{,}000--30{,}000 time frames per channel, which represents a few seconds of a discharge at 500\,kHz. The pipeline processes such a shot in approximately 0.5\,s on a single NVIDIA V100 \gls{gpu} and 5--10\,s on a \gls{cpu} depending on the number of channels processed in parallel, supporting between-shot monitoring as well as offline batch analysis. Per-domain metrics are summarized in Table~\ref{tab:domain_summary}.

\begin{table}[t]
    \centering
    \caption{Per-domain evaluation. Recall is at the default threshold (0.5) for TJ-II and DCLDE and at the validation-selected operating point for RadDet. F1 is reported at the F1-maximizing operating point; RadDet settings are selected on validation data only. PR-AUC$_{\pm2}$ is the threshold-independent bin-level area under the precision--recall curve at a $\pm2$-bin tolerance (TJ-II and DCLDE only). Event AP is per-contour average precision for DCLDE and class-agnostic point-in-box event average precision for RadDet. RadDet comprises 40{,}002 unique test frames rendered at three resolutions (120{,}006 inputs).}
    \label{tab:domain_summary}
    \footnotesize
    \setlength{\tabcolsep}{3pt}
    \begin{tabular}{llccccc}
        \toprule
        Domain & Dataset & $n$ & Recall & PR-AUC$_{\pm2}$ & Event AP \\
        \midrule
        Fusion (tokamak) & DIII-D & 11  & --- & --- & --- \\
        Fusion (stellarator) & TJ-II 2021 & 493 & 0.825 & 0.82 & --- \\
        Bioacoustics & DCLDE 2011 & 5000 & 0.72 & 0.73 & --- \\
        Radar & RadDet & 40{,}002 & 0.84 & --- & 0.82 \\
        \bottomrule
    \end{tabular}
\end{table}

\section{Discussion}\label{sec:discussion}
In many areas of research, the rate at which sensors produce time-frequency data outpaces the rate at which expert annotators can label them. For example, in fusion, experimental campaigns are anticipated to generate terabytes of multichannel fluctuation data per discharge \cite{churchillFrameworkInternationalCollaboration2021a}. Within these measurements, events of interest are sparse, structurally diverse, and unlabeled at the scale needed to train supervised models from scratch \cite{chenShortLargeScaleHeterogeneous2026}. This can introduce annotation coverage bias toward recognized event classes, leaving unfamiliar but potentially important structures absent from supervised label sets. The results here support a route around this bottleneck for the fluctuation-extraction subproblem.

We have shown that a surrogate model trained only on DIII-D can work to extract events in one step. Additionally, this model is general enough to transfer zero shot to a diverse field of data, including stellarator diagnostics, marine bioacoustic recordings, and wideband synthetic radar benchmark datasets. The data pipeline combines three optimization-based stages: penalized broadband baseline estimation which separates the vertical bias from time-frequency signals, complex-valued multichannel denoising trained to minimize cross-channel reconstruction error, and knee-point thresholding that maximizes the perpendicular distance between the normalized CDF and its endpoint chord. This transfer result is evidence that useful time-frequency structure is shared by measurements from instruments as different as laser interferometers and hydrophones.

The pipeline's nominal false positives track signal instead of than background (Section~\ref{sec:results-fp}). This is consistent with incomplete annotation coverage, and on \gls{dclde} intensity and morphology provide two measurements pointing in that direction (Supplementary Table~\ref{tab:dclde_fp_morphology}).

The class of signals targeted by this pipeline extends well beyond the domains evaluated here. Plausible domains include seismology, electroencephalography \cite{grandchampSingleTrialNormalizationEventRelated2011}, passive sonar, vibration-based fault detection, and radio astronomy.

Future work could investigate transfer learning from other domains instead of only tokamak signals. For tokamak data, future work could incorporate phase and turbulence information as additional pipeline stages, extend the denoiser to extract inherent cross-diagnostic correlation in a single pass, and pursue quantization and pruning for low-power \gls{cpu}-side deployment in continuous-monitoring contexts. Architecturally, the current multichannel denoising step can be made more memory efficient as it currently predicts each channel from the others one at a time. A blind-spot architecture applied across channels rather than pixels (cf.\ \cite{laineHighQualitySelfSupervisedDeep2019}) could address this.

\section{Materials and Methods}\label{sec:methods}
\subsection{Study objectives and design}\label{sec:study-design}

The study was designed to test whether a coherent-event extractor learned without manual labels from heterogeneous DIII-D diagnostics could transfer across instruments and scientific domains without updating model parameters. The workflow comprised four stages: complex time-frequency transformation, broadband baseline correction, cross-channel self-supervised denoising, and knee-point thresholding, followed by training a single-channel surrogate on the resulting targets. DIII-D observations were used for source-domain training, and the resulting surrogate was evaluated on held-out DIII-D observations and zero shot on TJ-II stellarator magnetics, DCLDE 2011 marine bioacoustics, and RadDet wideband radar. The evaluation protocol and dataset-specific processing are described below.

\subsection{Taxonomy of fluctuating signals}\label{sec:taxonomy}

In order to generalize a procedure across observation types, we first define the characteristics of the signals targeted by this pipeline. We use an operational taxonomy based on the deterministic versus random and stationary versus non-stationary structure of stochastic processes \cite{bendatRandomDataAnalysis2000}. Signals are categorized into five working classes: coherent, quasi-coherent, transient, broad, and stochastic. For the sequence of processing operations used here, these are grouped into three super-categories: coherent measurements (coherent + quasi-coherent), broadband measurements (transient + broad), and noise (stochastic). The grouping is summarized in Fig.~\ref{fig:concept}.

Deterministic periodic behavior arises from resonant processes and presents as narrow-band frequency content that persists over time and may chirp as energy distributions evolve. In tokamaks, this includes \glspl{ae}, \glspl{tm}, and geodesic acoustic modes \cite{kimMHDModeIdentification1999, chenPhysicsAlfvenWaves2016}; analogous narrow-band structures appear as dolphin tonal whistles in marine bioacoustics and as linear-frequency-modulated chirps in radar \cite{janikSignatureWhistleVariations1994, klauderTheoryDesignChirp1960}. Non-periodic deterministic behavior may manifest as transient bursts like \glspl{elm}, radar pulses, and echolocation click bursts in cetacean recordings, presenting in the time-frequency domain as impulses with broadband spectral signatures \cite{federiciKeyITERPlasma2003,versluisHowSnappingShrimp2000}.

Stochastic processes contribute zero-mean noise that may be Gaussian, Laplacian, or uniform depending on the sensor's noise physics. Non-stationary cascades follow approximate $1/f^{\chi}$ dropoffs with $0 < \chi \le 2$ that characterize phenomena including turbulence and integrated Brownian processes \cite{milotti1NoisePedagogical2002, schroederFractalsChaosPower2009}; their broad time-frequency spread motivates grouping them with transient events under the broadband category. The quasi-coherent class accommodates signals between pure oscillation, stochastic chaos, and broadband cascade, including modes well-documented in tokamak literature \cite{mazurenkoExperimentalTheoreticalStudy2002} and analogous staggered or chirping structures in other domains.

This categorization fixes the order of operations for extracting a signal: broadband measurements are isolated first since their high-amplitude structure would otherwise mask faint coherent modes, the residual coherent-plus-noise spectrogram is then denoised across multiple sensors, and candidate events are finally isolated by thresholding.

\subsection{Label-free event extraction}\label{sec:recipe}

The extraction pipeline is four sequential transformations that take a multichannel time-frequency tensor to a per-event segmentation mask, requiring no manual labels.

\subsubsection{Time-frequency representation}
We use the \gls{stft} for its computational efficiency and compatibility with existing fusion workflows \cite{straitMagneticDiagnosticSystem2006, zapata-cornejoNovelUnsupervisedMachine2024}, though the downstream pipeline is agnostic to the choice of time-frequency representation. Each spectral domain is processed to its own native frequency band (Section~\ref{sec:evaluation}) \cite{harrisUseWindowsHarmonic1978, chenShortLargeScaleHeterogeneous2026}.

The complex \gls{stft} $Z(t,f)$ decomposes additively into three components corresponding to the taxonomy:
\begin{equation}
Z(t,f) = M(t,f) + V(t,f) + \eta(t,f)
\end{equation}
where $M$ is the coherent component, $V$ the broadband component, and $\eta$ stochastic noise. The pipeline targets sequential estimation of each component. The processing pipeline is demonstrated on an \gls{ece} channel from shot 178631 in Fig.~\ref{fig:concept}. Part of this pipeline is implemented in the \gls{faith} Python package \cite{jalalvandFusionAIToolkit2026}.

\subsubsection{Broadband-coherent separation}
Sensors that resolve non-stationary phenomena often produce broadband content following an approximate $1/f^{\chi}$ power-law cascade that can obscure faint coherent modes. Existing practices in tokamak diagnostic analysis often handle this by clipping low-frequency bins below a fixed cutoff \cite{vanzeelandIsotopeImpactAlfven2024}, but this discards potentially useful information. The problem resembles harmonic-percussive separation in music \cite{fitzgeraldHarmonicPercussiveSeparation2010a}, but the broadband component cannot be cleanly modeled by a single digital filter. We instead adapt asymmetric least-squares baseline removal from event-related spectral perturbation analysis in \gls{eeg} \cite{grandchampSingleTrialNormalizationEventRelated2011, baekBaselineCorrectionUsing2014}. Writing $Z_R(t,f)=\operatorname{Re}Z(t,f)$ and $Z_I(t,f)=\operatorname{Im}Z(t,f)$, we fit the two component spectrograms independently. For each $q\in\{R,I\}$, the broadband baseline $V_q(t,f)$ is estimated by solving
\begin{equation}\label{eq:baseline}
\min_{V_q} \sum_{t,f} w_q(t,f) \, \bigl(Z_q(t,f) - V_q(t,f)\bigr)^2 + \lambda \sum_{t,f} \left(\nabla^2 V_q(t,f)\right)^2, \qquad q\in\{R,I\}.
\end{equation}
Here $w_q(t,f)$ is an asymmetric weight that down-weights positive residuals and $\lambda = 10^6$ regularizes baseline curvature. The weighted least-squares problems are solved independently by iteratively reweighted least squares: the weights are initialized to unity (an ordinary smoothness-penalized fit) and updated each iteration to down-weight bins lying above the current component baseline, until the weights stabilize \cite{baekBaselineCorrectionUsing2014}. Subtracting $V_q$ from $Z_q$ separately removes the frequency-dependent broadband intensity profile from each component and centers both residuals around a more uniform baseline across frequency. The resulting two-plane residual has an equalized background, enabling the subsequent cross-channel denoiser to operate effectively across the full band. Edge effects from imperfect baseline fitting are mitigated by an inverse-power-law pre-emphasis filter described in Supplementary Methods \cite{atalSpeechAnalysisSynthesis}. Broadband and transient events, which share spectral support, can be distinguished after separation by an impulse-amplitude measurement on filtered time slices if needed.

\subsubsection{Multi-channel self-supervised denoising}
Classical denoisers like Wiener filtering or BM3D tend to smear faint transients; sparse transforms (wavelet, curvelet) impose morphological priors; modal decompositions (empirical or variational mode decomposition) are computationally prohibitive at the scale required \cite{dabovImageDenoisingSparse2007, dragomiretskiyVariationalModeDecomposition2014}.

Multichannel methods are standard in fusion analysis but classical extensions such as singular spectrum analysis or the MUltiple SIgnal Classification (MUSIC) algorithm scale poorly with channel count \cite{grothMultivariateSingularSpectrum2011, olofssonArrayMagneticsModal2014a}. Hardware-level multichannel denoising such as correlation electron cyclotron emission \cite{whiteCorrelationElectronCyclotron2008} addresses the bias--variance tradeoff but is not always feasible and is not currently planned for next-generation devices.

Classical cross-spectral analysis starts from the \gls{cps},
\begin{equation}\label{eq:cps}
S_{xy}(f) = \mathbb{E}\!\left\{X_x(f)\, X_y(f)^{*}\right\}
\end{equation}
which estimates cross-channel covariance while incoherent random-phase components cancel under the expectation \cite{bendatRandomDataAnalysis2000}.  Under jointly proper complex Gaussian and stationary assumptions, the conditional mean has the linear form
\begin{equation}\label{eq:condexp}
\hat{X}_i = \mathbb{E}[X_i \mid \mathbf{X}_{-i}]
= \mathbf{s}_{i,-i}(f)\,\mathbf{S}_{-i,-i}(f)^{-1}\mathbf{X}_{-i},
\end{equation}
where $\mathbf{X}_{-i}$ stacks the conditioning channels, $\mathbf{S}_{-i,-i}$ contains their auto- and cross-power spectral densities, and $\mathbf{s}_{i,-i}$ contains the cross-power spectral densities between the target and conditioning channels. For two channels this reduces to $(S_{ij}/S_{jj})X_j$, so the \gls{cps} appears as the numerator of the classical transfer-function estimate. This is the standard linear minimum-mean-square-error estimator expressed through auto- and cross-spectral densities. A network trained under squared-error loss learns the conditional expectation directly, agreeing with this form in the linear Gaussian limit without restricting the conditional mapping to be linear or Gaussian (derivation in Supplementary Methods). Because the learned estimator need not average its reconstructed output over unrelated time segments, rare transients can be preserved wherever other channels contain information about them rather than being diluted in proportion to their duty cycle.

A U-Net is trained to approximate this expectation by mapping the stacked real-and-imaginary \glspl{stft} of all channels except the target ($2(k-1)$ planes, for $k$ channels) to the two-channel real-and-imaginary pair of the target channel, with \gls{mse} loss. While most spectrogram-based machine learning uses magnitude only, or magnitude and phase as two separate channels, we operate on the real and imaginary parts because they are the representation most amenable to self-supervised denoising: their noise is closer to zero-mean, whereas magnitude is positively biased under the Rayleigh distribution and phase wraps modulo $2\pi$ \cite{rayleighXIIResultantLarge1880}. The cross-product expands as $X_x X_y^* = (R_x R_y + I_x I_y) + i(I_x R_y - R_x I_y)$, so the U-Net operates on the same real-and-imaginary variables that define the classical cross-spectral covariance (Supplementary Fig.~\ref{fig:ablation_repr}). Self-supervised denoising theory supports this, provided the noise is zero-mean and independent across channels \cite{lehtinenNoise2NoiseLearningImage2018}. Information transfer of this kind across channels and modalities has been previously demonstrated for diagnostic reconstruction in tokamak plasmas \cite{jalalvandMultimodalSuperresolutionDiscovering2025}. Denoiser training is halted by early stopping when the \gls{tv} of denoised validation outputs stops decreasing (patience of three epochs); the rationale is given in Supplementary Methods.

\subsubsection{Parameter-free thresholding}
With coherent modes isolated, candidate events are extracted by thresholding the cleaned spectrogram. Otsu's method \cite{otsuThresholdSelectionMethod1979} assumes a bimodal intensity distribution and can perform poorly on the sparse, heavy-tailed distributions characteristic of spectrograms. Quantile-based thresholds work in practice but require a hand-chosen percentile. We instead apply knee-point detection on the \gls{cdf} of time-frequency bin intensities where the corner of the noise distribution is automatically selected as the threshold \cite{satopaaFindingKneedleHaystack2011}. The \gls{cdf} is computed on the upper half of intensities (with values below the spectrogram mean clamped to the mean to handle the long noise tail), normalized so its axes are equal, and the maximum perpendicular distance from the curve to the line connecting its endpoints is taken as the knee. The procedure is parameter-free in that no explicit threshold value is required; its only internal setting, the interpolation density used to trace the \gls{cdf}, does not meaningfully change the selected threshold. The construction was inspired by the observation in bioacoustic vocalization extraction that true positives concentrate above the \gls{cdf} knee \cite{dadouchiAutomatedSegmentationLinear2013}. However, this requires that a spectrogram must contain at least one signal since otherwise the curve may select arbitrary background points and return what practitioners may attribute a false positive. Thus, all 23 training shots in this work were chosen only if they contained observable signal in at least one of the diagnostics.

\subsection{Surrogate model training}\label{sec:surrogate}

The pipeline to this point yields a database of approximately 40{,}000 segmented event masks from the source tokamak data without any manual labels. To improve coverage of sensor-specific events that the cross-channel denoiser cannot reliably infer from other channels, owing to the sparsity of such events in the data, we train an ensemble of five refinement U-Nets via 5-fold cross-validation (one per held-out fold) with symmetric binary cross-entropy loss on single-channel inputs and standard image-segmentation augmentations including flips, and elastic deformation \cite{wangSymmetricCrossEntropy2019, ronnebergerUNetConvolutionalNetworks2015}. Pixels where this refinement ensemble assigns appreciable probability (mean sigmoid probability above 0.25) but is not yet saturated (predictive binary entropy above 0.45\,nats), and that do not conflict with the opposite class, are added to the candidate label set as pseudo-labels \cite{liUsingDeepLearning2023}; all existing labels are retained. This expands coverage into plausible signals that the current labels miss without relabeling confidently decided regions.

A final surrogate U-Net is then trained on this labeled set with spectrogram-targeted augmentations including SpecAugment \cite{parkSpecAugmentSimpleData2019}. Per-sensor running mean and standard deviation are used for normalization, with robust scaling clipped at the 0.1 and 99.9 percentiles \cite{maronnaRobustStatisticsTheory2019}.

A single U-Net architecture is used across all three roles in the pipeline: a multichannel denoiser, single-channel refiner, and final surrogate. At inference only the final surrogate is applied; the refinement ensemble serves solely to generate training pseudo-labels. The U-Net follows the original construction \cite{ronnebergerUNetConvolutionalNetworks2015} with transpose convolutions replaced by upsample-double-convolution to suppress checkerboard artifacts \cite{odenaDeconvolutionCheckerboardArtifacts2016} (Extended Data Fig.~\ref{fig:model}). The architecture is deliberately generic since this work aims to lay out a base pipeline based on the correspondence between self-supervised denoising and classical signal enhancement. Constructing the training database from 5{,}000 raw DIII-D shot segments takes approximately 5 hours on a single NVIDIA A100 \gls{gpu} with a 64-core \gls{cpu}, and 5-fold training of the final surrogate takes approximately 12 hours on the same hardware.

\subsection{Cross-domain evaluation protocol}\label{sec:evaluation}

A surrogate model trained on DIII-D fluctuation data is evaluated zero-shot on four datasets spanning fusion plasmas, marine bioacoustics, and wideband radar benchmarks. Here zero-shot transfer means that no model parameters are updated in a target domain; it does not mean that the complete measurement pipeline is domain independent. Spectrogram parameterization is chosen for each domain. Results at the fixed default threshold of 0.5 require no target-label threshold calibration and are reported where available, while PR-AUC and AP summarize confidence sweeps. For \gls{raddet}, event post-processing and the F1 operating point are selected on one half of its validation split, confirmed on the untouched half, and frozen before the test split is evaluated. F1-optimal thresholds selected on annotated evaluation data for the other target datasets are reported separately from uncalibrated fixed-threshold and threshold-swept results.

\subsubsection{Evaluation metrics}\label{sec:eval-metrics}
Predictions are scored against bin-level reference masks. At exact match a predicted positive is a true positive only where the reference bin is also positive. Because time-frequency boundaries are intrinsically fuzzy due to spectral leakage or annotator variability, we look at relaxed metrics with a matching tolerance of $\tau$ bins, following established boundary-aware segmentation evaluation and problem-aware metric selection \cite{csurkaWhatGoodEvaluation2013, maier-heinMetricsReloadedRecommendations2024}. A prediction counts toward precision if a reference positive lies within $\tau$ bins (Chebyshev distance, i.e.\ within the $(2\tau+1)\times(2\tau+1)$ neighborhood obtained by dilating the reference mask by $\tau$ bins), and a reference positive counts toward recall if a prediction exceeds threshold anywhere in that same $\tau$-bin neighborhood. We report $\tau \in \{0,1,2,3\}$. We use $\tau=2$ as a representative relaxed match because it captures one- to two-bin boundary discrepancies while retaining stricter and looser tolerances for comparison. Threshold sensitivity is summarized by the area under the precision-recall curve (PR-AUC), computed as the average precision over the swept detection threshold; PR-AUC is threshold-independent and so avoids the bias of the F1-optimal operating point. We also report the area under the receiver-operating-characteristic curve (ROC-AUC) for reference but do not rely on it since annotated mode bins are a small minority of time-frequency bins, which makes ROC-AUC overly optimistic \cite{saitoPrecisionRecallPlotMore2015}.

Because TokEye outputs masks instead of outputting boxes, our primary \gls{raddet} measure is task-specific point localization rather than the benchmark's standard IoU-based box detection \cite{huangRadDetWidebandDataset2025a}. A detection is correct when its highest-probability point falls inside an annotated event box, with one-to-one matching between detections and annotations. Event AP summarizes the resulting precision--recall curve across confidence thresholds.

\subsubsection{DIII-D tokamak dataset}
Training data for DIII-D signals come from four diagnostic types (magnetics, \gls{ece} radiometer, \gls{co2} interferometer, \gls{bes}) across 23 shots (161172, 161403, 163518, 166419, 170660, 170670, 170672, 170677, 170679, 170796, 175987, 176054, 178631, 180634, 184847, 184859, 184902, 184964, 190904, 193273, 193277, 193280, 193281). Shots used for evaluation are 170008, 185781, 199597, and 199607. Consistent with previous spectrogram-related tasks for DIII-D, all spectrograms are resampled to 500\,kHz using 8th-order Chebyshev Type~I decimation with phase preservation, then processed with a 2\,ms Hann window (1024 samples at 500\,kHz, 87.5\% overlap), resulting in spectrograms that span 0--250\,kHz. There is no bin-level ground truth for this set, so reported results are qualitative demonstrations of cross-diagnostic transfer within the source tokamak.

\subsubsection{TJ-II stellarator dataset}
The TJ-II dataset comprises 493 spectrogram images from TJ-II stellarator Mirnov-coil signals with hand-labeled binary masks of fluctuation modes in the 1--500\,kHz band used by Bustos et~al.\ \cite{bustosAutomaticIdentificationMHD2021}. The surrogate model trained on DIII-D is applied zero-shot to the dataset.

\subsubsection{DCLDE 2011 bioacoustic dataset}
The \gls{dclde} 2011 benchmark provides annotated marine-mammal vocalizations including: \emph{Delphinus capensis} (long-beaked common dolphin), \emph{Delphinus delphis} (short-beaked common dolphin), \emph{Stenella longirostris} (spinner dolphin), and \emph{Tursiops truncatus} (bottlenose dolphin, Southern California subset) \cite{rochAutomatedExtractionOdontocete2011}. Audio recordings (192\,kHz sample rate) are processed into log-magnitude spectrograms (8\,ms frames, 2\,ms step, 75\% overlap, clipped to the 5--50\,kHz band), then sliced into non-overlapping 250$\times$250 patches. Per-patch outputs are evaluated against rasterized contour annotations.

\subsubsection{RadDet radar dataset}
\gls{raddet} \cite{huangRadDetWidebandDataset2025a} is a public wideband radar pulse detection and modulation classification benchmark with synthetic spectrograms drawn from a 500\,MHz band and YOLO-format bounding boxes across 11 modulation classes. We evaluate all six dataset variants, combining three time-frequency resolutions (128$\times$128, 256$\times$256, 512$\times$512) with two signal-density environments (HW001 and HW009, with one and up to nine signals per frame, respectively). Each variant contains 6{,}001 validation and 20{,}001 test images. Within each density environment the three resolutions are renderings of the same underlying frames, giving 40{,}002 unique test frames, 120{,}006 resolution-specific inputs, and 100{,}066 unique radar events across the two environments.

Spectrograms are bilinearly resampled to an effective $512\times512$ before inference. The pixelwise maximum of the two output-channel probabilities is averaged between the original input and a time-reversed test-time augmentation, then mapped back to the native grid. Event post-processing is selected by maximizing event F1 on the even-indexed half of the validation split; the odd-indexed half is used only for confirmation, and test data are scored once after freezing the configuration. Five variants use Gaussian smoothing ($\sigma=1$), while 128HW009 uses a $3\times3$ grayscale opening. In table order, the selected candidate-score and merge-gap-fraction pairs are soft area/0.40, top-32 mean/0.10, local contrast/0.50, local contrast/0.10, top-decile mean/0.40, and maximum/0.10. No band-limiting or candidate confidence cutoff is applied. For a secondary coarse-overlap diagnostic, each grouped component is reduced to its energetic core, converted to a bounding extent, and scored at IoU 0.25; this diagnostic is not used to claim tight box estimation. Per-variant thresholds and results are reported in Supplementary Table~\ref{tab:raddet_per_variant}.

\backmatter

\bibliography{tokeye,manual-refs}

\bmhead{Acknowledgments}
Special thanks to Andres Bustos and Enrique Zapata-Cornejo for providing access to the TJ-II dataset. Thank you to Jalal Butt, Sangkyeun Kim, Ricardo Shousha, Hiro Farre, Sui Xiong Tay, Sui Jiet Tay, and Frederik Simons for valuable feedback on this work.

\noindent\textbf{Funding:}
This material is based upon work supported by the U.S. Department of Energy, Office of Science, Office of Fusion Energy Sciences, using the DIII-D National Fusion Facility, a DOE Office of Science user facility, under Awards DE-FC02-04ER54698 and DE-SC0024527. The authors also gratefully acknowledge financial support from the Princeton Laboratory for Artificial Intelligence under Award 2025-97.

Disclaimer: This report was prepared as an account of work sponsored by an agency of the United States Government. Neither the United States Government nor any agency thereof, nor any of their employees, makes any warranty, express or implied, or assumes any legal liability or responsibility for the accuracy, completeness, or usefulness of any information, apparatus, product, or process disclosed, or represents that its use would not infringe privately owned rights. Reference herein to any specific commercial product, process, or service by trade name, trademark, manufacturer, or otherwise does not necessarily constitute or imply its endorsement, recommendation, or favoring by the United States Government or any agency thereof. The views and opinions of authors expressed herein do not necessarily state or reflect those of the United States Government or any agency thereof.

\noindent\textbf{Author contributions:}
N.C. wrote the manuscript with input from all authors.

\noindent\textbf{Competing interests:}
The authors declare no competing interests.

\noindent\textbf{Data, code, and materials availability:}
All relevant DIII-D data supporting the findings of this study are available from the DIII-D National Fusion Facility, which is operated by General Atomics for the U.S.\ Department of Energy. Access to DIII-D data requires following the user protocols described on the \href{https://d3dfusion.org/become-a-user/}{DIII-D website}. Specific questions regarding data availability can be directed to the corresponding author of this paper. The processed source data underlying the figures and the scripts used to reproduce them will be deposited in Princeton Data Commons; during review they are available from the corresponding author.

TJ-II data are available on request of the authors of \cite{bustosAutomaticIdentificationMHD2021}, subject to the respective program data-sharing policies. \gls{dclde} 2011 is publicly available \cite{rochDCLDE2011Data2025}. \gls{raddet} is publicly available via the original paper \cite{huangRadDetWidebandDataset2025a}.

The full extraction pipeline and training code are openly available at \url{https://github.com/PlasmaControl/tokeye}\cite{Chen_PlasmaControl_tokeye_Release_for_2026}. The pretrained DIII-D surrogate weights are available at \url{https://huggingface.co/collections/nc1/tokeye}. This study did not generate new materials.

\clearpage
\setcounter{figure}{0}
\setcounter{table}{0}
\renewcommand{\theHfigure}{S\arabic{figure}}
\renewcommand{\theHtable}{S\arabic{table}}
\renewcommand{\figurename}{Supplementary Figure}
\renewcommand{\tablename}{Supplementary Table}

\section*{Supplementary Methods}\label{sec:supp_methods}

\subsection*{Baseline removal}

In the asymmetric least-squares fit of Equation~\ref{eq:baseline}, the baseline curvature may spike near the low-frequency edge due to imperfect fitting. Before fitting we therefore pre-emphasize the power spectrogram with an inverse-power-law weighting, multiplying $P(t,f)$ by a factor proportional to $f^{\chi}$, with $\chi$ a fixed constant rather than a per-channel fit, so that the $1/f^{\chi}$ cascade is approximately flattened, in analogy with the pre-emphasis filters of classical speech analysis \cite{atalSpeechAnalysisSynthesis}.

\subsection*{Comparison to linear complex multichannel denoising}

For a shot with $k$ usable sensor channels, e.g.\ individual magnetic coils, the denoiser input is the real-and-imaginary stack $\mathbf{X} \in \mathbb{R}^{2k \times F \times T}$ of the $k$ baseline-corrected \glspl{stft}, with the two planes of the target channel removed; the output is the predicted real-and-imaginary pair $\hat{\mathbf{Y}} \in \mathbb{R}^{2 \times F \times T}$ for that channel. At a fixed time-frequency bin each channel decomposes as $X_j = M_j + \eta_j$, with coherent component $M_j$ and zero-mean noise $\eta_j$ that is independent across channels; for the classical special case we take the channels jointly proper complex Gaussian and wide-sense stationary. Treating each channel as a zero-mean complex random variable, the natural inner product between two channels is their cross-power, $\langle X_i, X_j\rangle = \mathbb{E}[X_i X_j^*] = S_{ij}(f)$. The normalized quantity $\gamma_{ij}(f) = S_{ij}/\sqrt{S_{ii}S_{jj}}$ has magnitude that measures linear coherence and phase that measures relative phase. Coherent structure is the component shared across channels, while channel-independent noise is orthogonal to every other channel.

Under these assumptions, the conditional mean of one channel given the others has the standard linear minimum-mean-square-error form,
\begin{equation}\label{eq:linear_mmse}
\mathbb{E}\!\left[X_i \mid \{X_j\}_{j \neq i}\right] \;=\; \mathbf{s}_{i,-i}(f)\, \mathbf{S}_{-i,-i}(f)^{-1}\, \mathbf{X}_{-i},
\end{equation}
where $\mathbf{X}_{-i}$ stacks the conditioning channels, $\mathbf{S}_{-i,-i}(f)$ is their matrix of auto- and cross-power spectral densities, and $\mathbf{s}_{i,-i}(f)$ is the row vector of cross-power spectral densities between the target channel and each conditioning channel. The cross-power spectral densities are complex-valued while only the diagonal auto-spectra of $\mathbf{S}_{-i,-i}(f)$ are real \cite{bendatRandomDataAnalysis2000}. For two channels it reduces to $\mathbb{E}[X_i \mid X_j] = \left(S_{ij}(f)/S_{jj}(f)\right) X_j$, the classical transfer-function estimate whose numerator is the \gls{cps} of Equation~\ref{eq:cps}. Because channel-independent noise is orthogonal to the conditioning channels, this estimate passes cross-channel-coherent structure while suppressing that noise.

\begin{figure}[t]
    \centering
    \includegraphics[width=\textwidth]{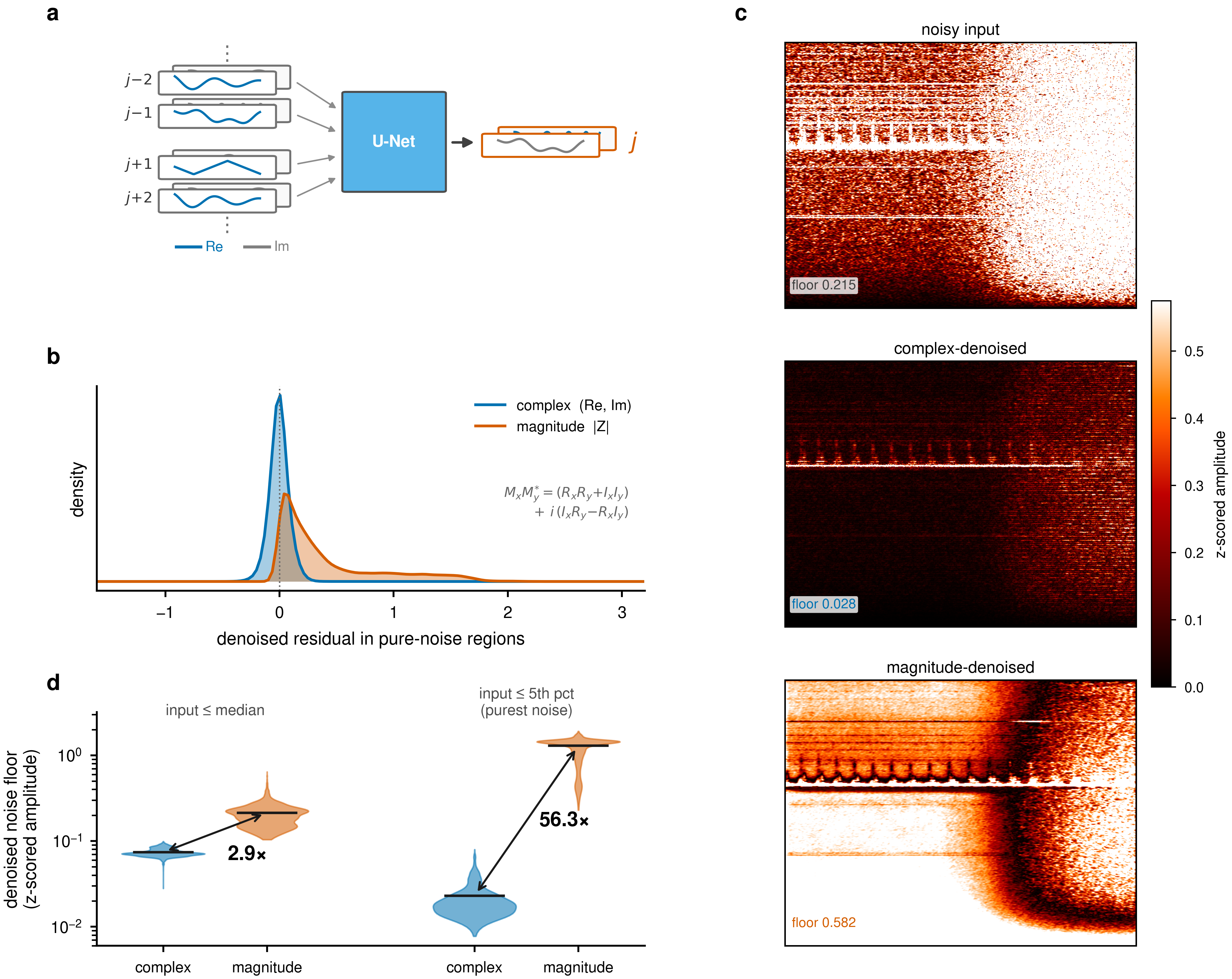}
    \caption{\textbf{Complex vs magnitude denoising.} (a)~A shared U-Net predicts each complex channel from the others. (b)~Denoised residual in pure-noise regions: approximately symmetric and near zero-mean for the complex pair, positively biased for magnitude. (c)~Example \gls{ece} spectrogram denoised under each representation. (d)~Residual noise floor across the 4{,}600 channel-maps: 2.9 times higher for magnitude at the median, 56.3 times on the lowest-amplitude bins (below the 5th input percentile).}
    \label{fig:ablation_repr}
\end{figure}

A network $g$ trained to predict one channel from the others learns the same conditional quantity without restricting the mapping to be linear, converging under \gls{mse} loss to the conditional expectation,
\begin{equation}
g^*\!\left(\{X_j\}_{j \neq i}\right) = \mathbb{E}\!\left[X_i \mid \{X_j\}_{j \neq i}\right] = \mathbb{E}\!\left[M_i \mid \{X_j\}_{j \neq i}\right],
\end{equation}
because $\eta_i$ is independent of every input channel and so contributes only its average. Cross-channel-coherent structure is preserved while channel-independent noise cannot be learned \cite{lehtinenNoise2NoiseLearningImage2018, jangSelfsupervisedImageDenoising2023}. The linear formula above is therefore a classical reference case, while the network can learn a nonlinear conditional mapping. Baseline correction removes the frequency-dependent broadband intensity profile and centers the residual around a common baseline across frequency, creating the spectrally equalized representation on which the denoiser operates. An empirical comparison (Supplementary Fig.~\ref{fig:ablation_repr}) measures how far each representation departs from this idealization on real diagnostics: the complex residual remains near zero while the magnitude residual retains a positive bias.

Single-image blind-spot methods, by contrast, mask individual pixels and reconstruct them from their spatial neighborhood, which presumes that noise is independent from pixel to pixel and that signal is locally redundant \cite{chenExploringEfficientAsymmetric2024}. The \gls{stft}'s overlapping windows correlate noise across neighboring bins, so a masked noisy bin can be reconstructed from its neighbors and the noise leaks through the blind spot; conversely the signal of interest occupies thin, sparse traces, so masking removes exactly the evidence needed to reconstruct them and the denoiser suppresses faint modes.

\subsection*{Total-variation stopping criterion}

Self-supervised denoising is expected to converge, but with a large enough model on a small enough dataset the denoiser may eventually begin to overfit to residual noise correlations. Its output \gls{tv} then rises, since noise is the highest-variation component of the spectrogram. We therefore monitor the \gls{tv} of denoised validation outputs during training and halt once it no longer decreases, with a patience of three epochs \cite{rudinNonlinearTotalVariation1992}.

\subsection*{Ablation design and reduced pipeline}

We use a leave-one-out ablation on the TJ-II 2021 dataset to isolate the three design choices behind the coherent-mode pipeline: the complex representation, broadband baseline removal, and multichannel self-supervised denoising. We separately measure the effect of complex versus magnitude on the denoiser's residual noise.

Each variant is trained with 5-fold cross-validation on an intermediate uncertainty-aware refinement network which uses Monte Carlo dropout plus a final surrogate U-Net. This model is evaluated on 493 TJ-II 2021 images. Per fold, we report recall at the fixed 0.5 threshold and F1 at the per-fold F1-optimal threshold. This ablation is self-contained and uses a reduced dataset. First, one tenth of the original training data is used, drawn from \gls{ece}, magnetics, and \gls{bes}. Second, it applies a simplified per-sample normalization. As a consequence the ablation should be read as a relative comparison of design choices, not as a measure of absolute performance.

\subsubsection*{Results}

\begin{table}[t]
\centering
\caption{5-fold reduced ablation on TJ-II 2021. F1 is at the per-fold F1-optimal threshold.}
\label{tab:ablation}
\setlength{\tabcolsep}{3pt}
\begin{tabular}{lcc}
\toprule
Variant & F1 @opt & F1-opt thr \\
\midrule
Full (complex)        & 0.303 $\pm$ 0.029 & 0.10 \\
Magnitude             & 0.319 $\pm$ 0.056 & 0.05 \\
No baseline removal   & 0.197 $\pm$ 0.019 & $\sim$0.08 \\
No denoising          & 0.108 $\pm$ 0.067 & 0.05 \\
\midrule
\textbf{Deployed model (reference)} & \textbf{0.478} & 0.80 \\
\bottomrule
\end{tabular}
\end{table}

\begin{table}[t]
\centering
\caption{Comparison of design choices on TJ-II (F1-optimal threshold; paired $t$-test, $n=5$ folds). Bold $p$-values account for Bonferroni correction over $m=4$ comparisons.}
\label{tab:ablation_stats}
\setlength{\tabcolsep}{6pt}
\begin{tabular}{lcc}
\toprule
Comparison (factor) & mean $\Delta$F1 & $p$ \\
\midrule
Full vs.\ no denoising (denoising)        & $+0.196$ & \textbf{0.0010} \\
Magnitude vs.\ no denoising (denoising)   & $+0.211$ & \textbf{0.0065} \\
Full vs.\ no baseline removal (baseline)  & $+0.106$ & \textbf{0.0070} \\
Full vs.\ Magnitude (representation)      & $-0.015$ & 0.63 \\
\bottomrule
\end{tabular}
\end{table}

Across folds, removing the cross-channel denoiser is the dominant degradation ($\Delta\mathrm{F1} = +0.196$, $p < 0.01$; Supplementary Table~\ref{tab:ablation_stats}); baseline removal contributes a smaller but significant amount ($+0.106$); and the complex-versus-magnitude representation is not significant at the downstream operating point ($p = 0.63$), its effect appearing instead upstream, in the denoiser noise floor (Supplementary Fig.~\ref{fig:ablation_repr}).

\section*{Supplementary Tables}

Per-domain quantitative detail referenced from Results is given in Supplementary Tables~\ref{tab:tjii_vs_bustos}--\ref{tab:raddet_per_variant}.

\begin{table}[t]
\centering
\caption{TJ-II head-to-head. TokEye is plasma-trained on DIII-D but evaluated zero-shot on the public TJ-II 2021 test set; Bustos et al.\ \cite{bustosAutomaticIdentificationMHD2021} is a U-Net trained directly on TJ-II with in-domain pixel labels. Bustos et al.\ do not report precision or F1.}
\label{tab:tjii_vs_bustos}
\begin{tabular}{lcccc}
\toprule
Method & thr & Recall & Precision & F1 \\
\midrule
Bustos 2021 (supervised) & --- & 0.579 & --- & --- \\
TokEye (zero-shot, default thr=0.5) & 0.50 & 0.825 & 0.170 & 0.283 \\
TokEye (zero-shot, F1-optimal thr=0.8) & 0.80 & 0.531 & 0.434 & 0.478 \\
\bottomrule
\end{tabular}
\end{table}

\begin{table}[t]
\centering
\caption{DCLDE 2011 \emph{event-level} zero-shot detection, pooled across the four dolphin species. A predicted connected component matches an annotated whistle under two contour-aware criteria: any-overlap ($\geq 1$ shared pixel) and center-hit (prediction passes through the contour centroid). Per-species results are given in Supplementary Table~\ref{tab:dclde_per_species}.}
\label{tab:dclde_event}
\begin{tabular}{lcccc}
\toprule
Criterion & Recall & Precision & F1 & AP \\
\midrule
any-overlap & \textbf{0.784} & 0.114 & 0.192 & \textbf{0.593} \\
center-hit  & 0.578 & 0.083 & 0.139 & 0.370 \\
\bottomrule
\end{tabular}
\end{table}

\begin{table}[t]
\centering
\caption{DCLDE 2011 zero-shot evaluation across four dolphin species, at the event and pixel level. Event-level detection uses the any-overlap criterion (Supplementary Table~\ref{tab:dclde_event}); pixel-level F1-optimal thresholds are selected per species from the PR sweep.}
\label{tab:dclde_per_species}
\small
\setlength{\tabcolsep}{2.5pt}
\begin{tabular}{lcccc|cccc}
\toprule
 & \multicolumn{4}{c|}{event-level (any-overlap)} & \multicolumn{4}{c}{pixel-level} \\
Species & Recall & Prec. & F1 & AP & Recall@0.5 & F1@0.5 & F1-opt thr & F1-opt \\
\midrule
\textit{D. capensis}          & 0.771 & 0.115 & 0.201 & 0.570 & 0.766 & 0.181 & 0.85 & \textbf{0.305} \\
\textit{D. delphis}           & 0.856 & 0.045 & 0.085 & 0.665 & 0.791 & 0.222 & 0.85 & \textbf{0.342} \\
\textit{S. longirostris}      & 0.790 & 0.096 & 0.171 & 0.593 & 0.693 & 0.229 & 0.85 & \textbf{0.300} \\
\textit{T. truncatus (SoCal)} & 0.718 & 0.200 & 0.312 & 0.545 & 0.626 & 0.212 & 0.80 & \textbf{0.271} \\
\midrule
mean & \textbf{0.784} & 0.114 & 0.192 & \textbf{0.593} & \textbf{0.719} & \textbf{0.211} & --- & \textbf{0.305} \\
\bottomrule
\end{tabular}
\end{table}

\begin{table}[t]
\centering
\caption{DCLDE false-positive morphology analysis. Each predicted connected component is assigned a morphology class --- tonal (narrowband, like a whistle), broadband (like a click or snap), or short (brief transient) --- and the class distribution is compared against the true-positive population.}
\label{tab:dclde_fp_morphology}
\begin{tabular}{lccc}
\toprule
Population & tonal & broadband & short \\
\midrule
True positive   & 0.74 & 0.22 & 0.04 \\
False positive  & 0.72 & 0.05 & 0.23 \\
\bottomrule
\end{tabular}
\end{table}

\begin{table}[t]
\centering
\caption{RadDet zero-shot evaluation across six dataset variants (resolution $\times$ signal density). Primary event metrics use class-agnostic one-to-one matching of each candidate's probability peak to an annotated event box. Post-processing and F1 operating thresholds are selected on one half of the validation split, checked on its untouched half, and frozen for test; no candidate confidence cutoff is applied. Coarse box AP@0.25 is included only as a lenient overlap diagnostic for mask-derived extents, not as a primary box-detection metric.}
\label{tab:raddet_per_variant}
\small
\begin{tabular}{lcccccc}
\toprule
Variant & Thr. & Event P & Event R & Event F1 & Event AP & Coarse box AP@0.25 \\
\midrule
128HW001 & 0.65 & 0.901 & 0.885 & 0.893 & 0.854 & 0.718 \\
128HW009 & 0.55 & 0.881 & 0.783 & 0.829 & 0.753 & 0.621 \\
256HW001 & 0.60 & 0.933 & 0.906 & 0.919 & 0.874 & 0.709 \\
256HW009 & 0.45 & 0.915 & 0.802 & 0.855 & 0.775 & 0.612 \\
512HW001 & 0.50 & 0.920 & 0.915 & 0.917 & 0.895 & 0.610 \\
512HW009 & 0.40 & 0.926 & 0.758 & 0.834 & 0.741 & 0.596 \\
\midrule
mean & --- & 0.913 & 0.841 & 0.874 & \textbf{0.815} & 0.644 \\
\bottomrule
\end{tabular}
\end{table}

\clearpage

\setcounter{figure}{0}
\setcounter{table}{0}
\renewcommand{\theHfigure}{ED\arabic{figure}}
\renewcommand{\theHtable}{ED\arabic{table}}
\renewcommand{\figurename}{Extended Data Fig.}
\renewcommand{\tablename}{Extended Data Table}

\section*{Extended Data}

\begin{figure}[t]
    \centering
    \includegraphics[width=\textwidth]{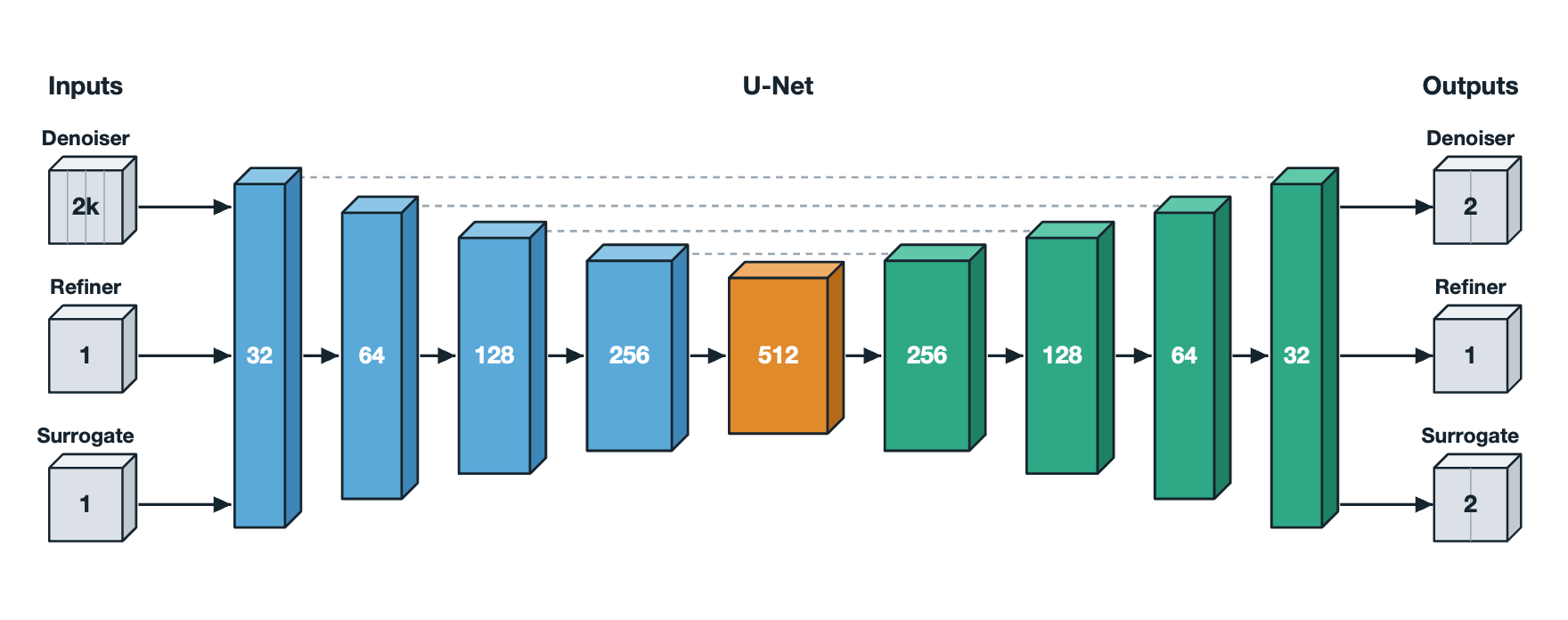}
    \caption{\textbf{U-Net architecture.} Given $k$ channels of an observation, the denoising step uses $2k$ inputs and $2$ outputs, the refining step uses $1$ input and $1$ output, and surrogate step uses $1$ input and $2$ outputs (coherent + transient detections).}
    \label{fig:model}
\end{figure}

\begin{figure}[t]
    \centering
    \includegraphics[width=\textwidth]{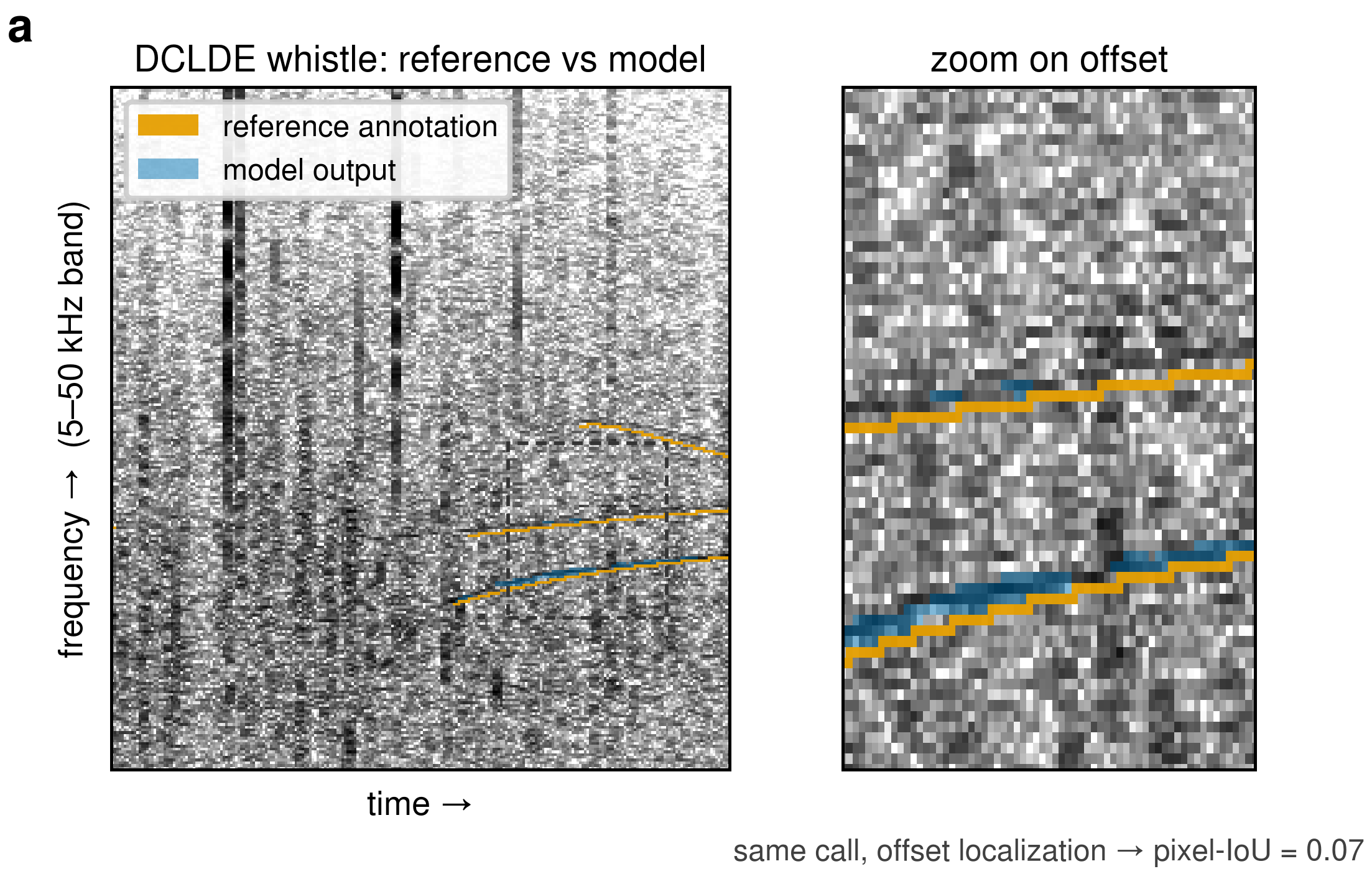}
    \caption{\textbf{\gls{dclde} failure mode.} The annotation marks the call's edge while the model localizes its center, so bin overlap is low despite detecting the same event.}
    \label{fig:dclde_failure}
\end{figure}

\begin{figure}[t]
    \centering
    \includegraphics[width=\textwidth]{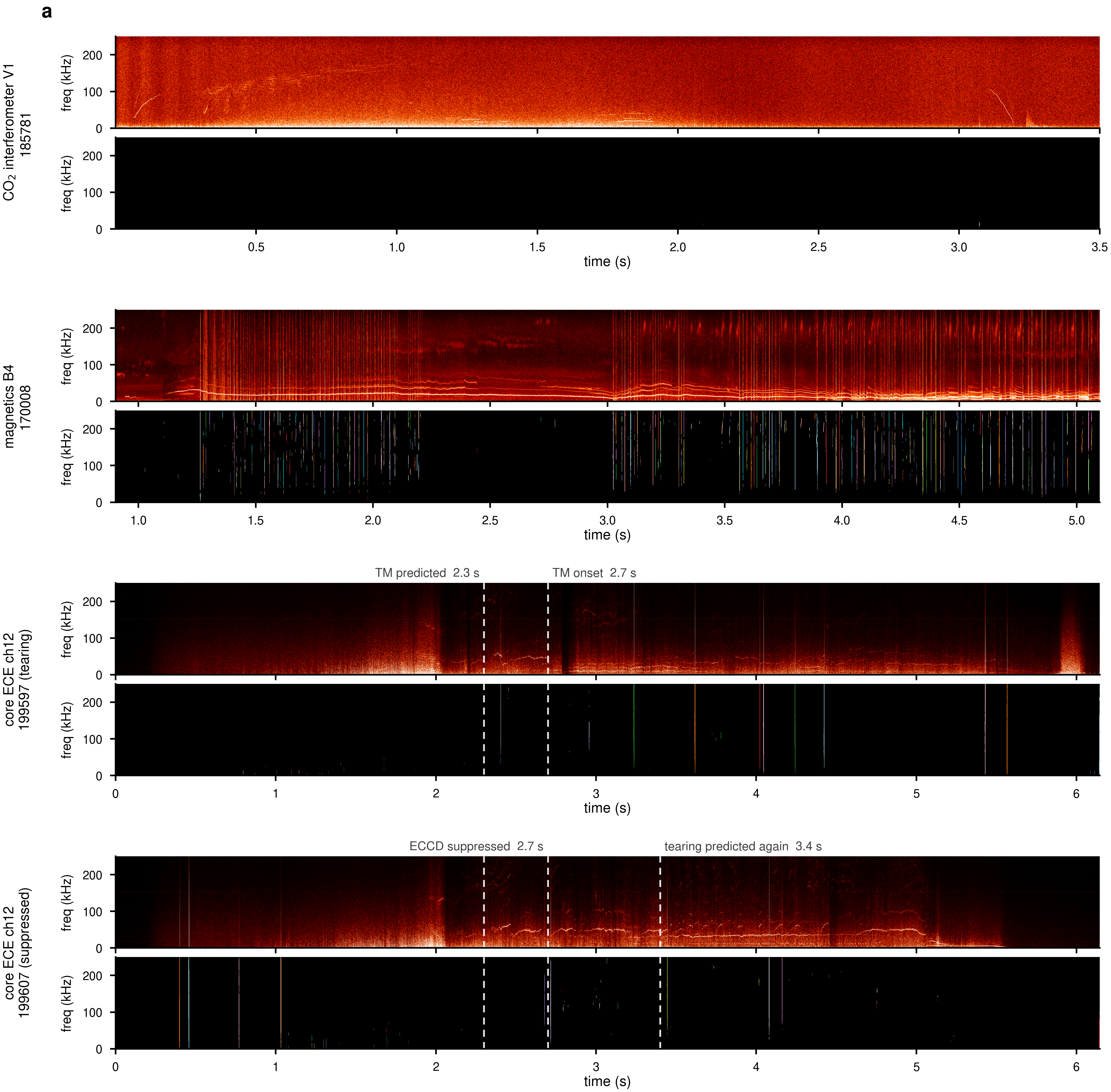}
    \caption{\textbf{Transient-channel output of the surrogate.} On the shots of Fig.~\ref{fig:diiid}, the second output channel (Extended Data Fig.~\ref{fig:model}) isolates broadband impulsive events as vertical streaks; the \gls{co2} channel, with no transient activity, stays empty. Shown for completeness as transient extraction is not the focus of this work.}
    \label{fig:diiid_transient}
\end{figure}

\begin{figure}[t]
    \centering
    \includegraphics[width=0.85\textwidth]{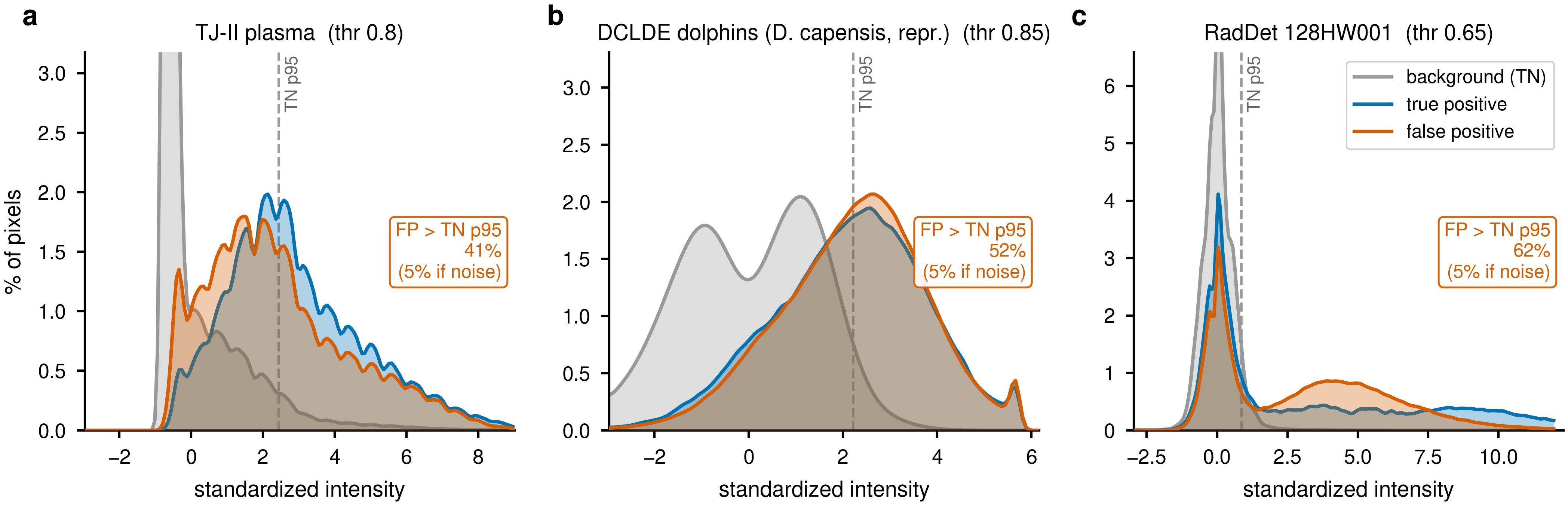}
    \caption{\textbf{Nominal false positives have signal-like intensity.} Intensity distributions of false-positive (FP), true-positive (TP), and background (TN) time-frequency bins at F1-maximizing thresholds selected from annotated evaluation data on TJ-II and a representative \glsentryshort{dclde} species (\textit{D.\ capensis}, whose FP-above-TN-p95 fraction of 0.52 equals the four-species mean), and from the held-out validation split for synthetic \glsentryshort{raddet} 128HW001. This complements the boundary-tolerance analysis of Fig.~\ref{fig:pr_curves}.}
    \label{fig:fp_intensity}
\end{figure}

The U-Net architecture (Extended Data Fig.~\ref{fig:model}), a \gls{dclde} failure-mode example (Extended Data Fig.~\ref{fig:dclde_failure}), the transient-channel output on the DIII-D evaluation shots (Extended Data Fig.~\ref{fig:diiid_transient}), and the false-positive intensity analysis (Extended Data Fig.~\ref{fig:fp_intensity}) are provided here.

\end{document}